\documentclass[journal,10pt]{IEEEtran}
\usepackage{cite}
\usepackage{balance}
\usepackage{color}
\usepackage{epsfig}
\usepackage{epstopdf}
\usepackage{graphicx}
\usepackage{tabularx}
\usepackage{booktabs}

\usepackage{varwidth}

\usepackage{amssymb,amsmath,amsthm}

\usepackage{bm}
\usepackage{algorithm}
\newfloat{routine}{htbp}{loa}
\floatname{routine}{Routine}
\makeatletter\newcommand\l@subroutine{\@dottedtocline{1}{1.5em}{2.3em}}\makeatother
\usepackage{algpseudocode}
\usepackage{pict2e}
\makeatletter
\def\BState{\State\hskip-\ALG@thistlm}

\makeatother

\usepackage{colortbl}

\begin{document}

\title{Collaborative Mobile Edge Computing in 5G Networks: New Paradigms, Scenarios, and Challenges}

\author{Tuyen~X.~Tran,~\IEEEmembership{Student Member,~IEEE,} Abolfazl Hajisami,~\IEEEmembership{Student Member,~IEEE,} Parul Pandey,~\IEEEmembership{Student Member,~IEEE,} and
Dario~Pompili,~\IEEEmembership{Senior Member,~IEEE}

\IEEEcompsocitemizethanks{\IEEEcompsocthanksitem The authors are with the Dept. of Electrical and Computer Engineering, Rutgers University--New Brunswick, NJ, USA.\protect\\
E-mails: \{tuyen.tran, hajisamik, parul$\_$pandey, pompili\}@cac.rutgers.edu\protect\\
\IEEEcompsocthanksitem This work was supported in part by the National Science Foundation (NSF) under Grant~No.~CNS-1319945.
}
}

\markboth{IEEE Communications Magazine, Special Issue on Fog Computing and Networking, April~2017}
{Shell \MakeLowercase{\textit{et al.}}: Bare Demo of IEEEtran.cls for Computer Society Journals}

\maketitle

\thispagestyle{empty}\pagenumbering{arabic}
\IEEEdisplaynotcompsoctitleabstractindextext
\IEEEpeerreviewmaketitle

\begin{abstract}
Mobile Edge Computing~(MEC) is an emerging paradigm that provides computing, storage, and networking resources within the edge of the mobile Radio Access Network~(RAN). MEC servers are deployed on generic computing platform within the RAN and allow for delay-sensitive and context-aware applications to be executed in close proximity to the end users. This paradigm alleviates the backhaul and core network and is crucial for enabling low-latency, high-bandwidth, and agile mobile services. This article envisages a real-time, context-aware collaboration framework that lies at the edge of the RAN, comprising MEC servers and mobile devices, and that amalgamates the heterogeneous resources at the edge. Specifically, we introduce and study three representative use-cases ranging from mobile-edge orchestration, collaborative caching and processing, and multi-layer interference cancellation. We demonstrate the promising benefits of the proposed approaches in facilitating the evolution to 5G networks. Finally, we discuss the key technical challenges and open-research issues that need to be addressed in order to make an efficient integration of MEC into 5G ecosystem.
\end{abstract}

\section{Introduction}
Over the last few years, our daily lifestyle is increasingly exposed to a plethora of mobile applications for entertainment, business, education, health care, social networking, etc. At the same time, mobile data traffic is predicted to continue doubling each year.  To keep up with these surging demands, network operators have to spend enormous efforts to improve users' experience, while keeping a healthy revenue growth. To overcome the limitations of current Radio Access Networks (RANs), the two emerging paradigms have been proposed: (i) Cloud Radio Access Network (C-RAN), which aims at the centralization of Base Station (BS) functions via virtualization, and (ii) Mobile Edge Computing~(MEC), which proposes to empower the network edge. While the two technologies propose to move computing capabilities to different direction (to the cloud versus to the edge), they are complementary and each has a unique position in the 5G ecosystem.

\begin{figure*}
 \centering
 \includegraphics[scale = .9]{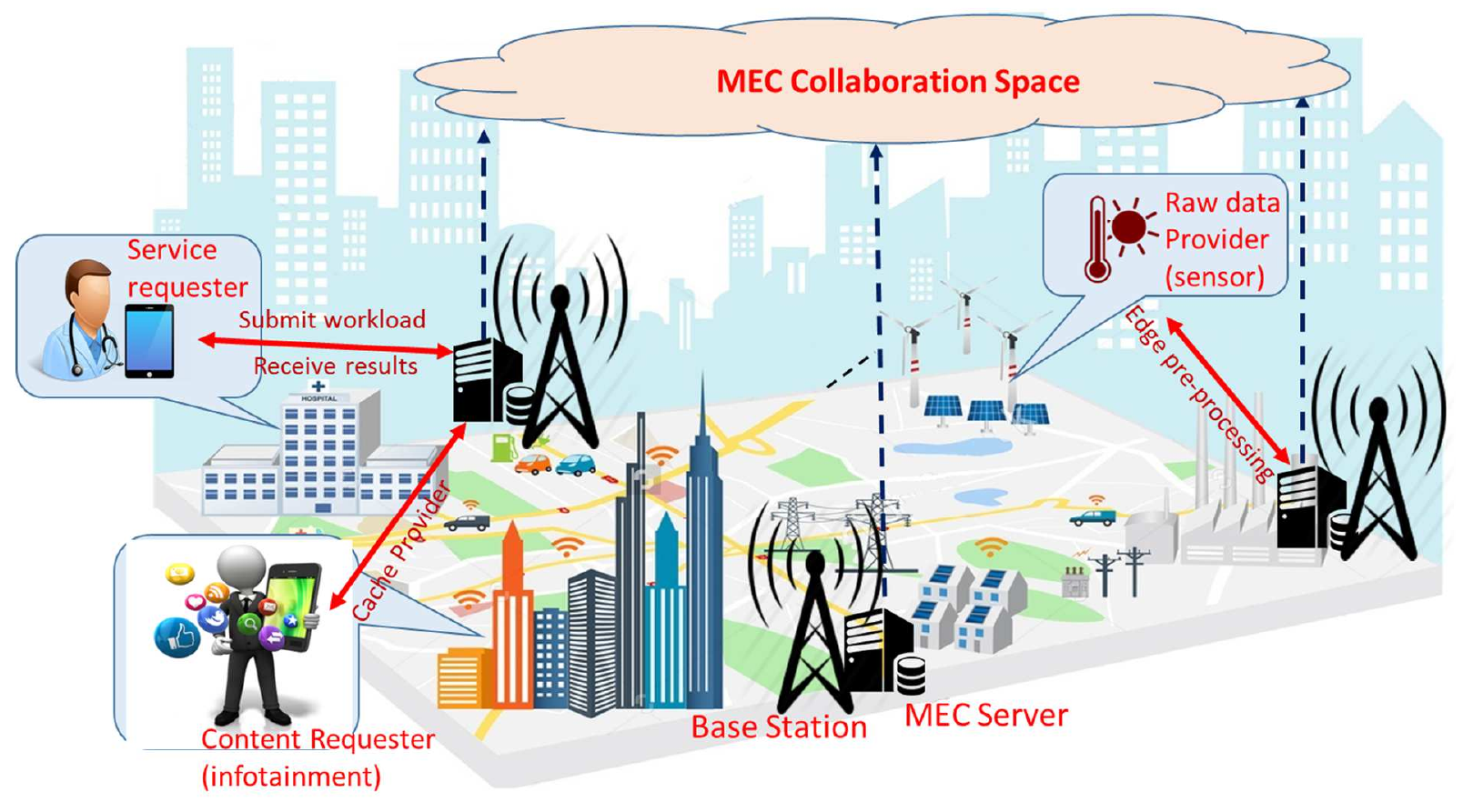}
\caption{ Illustration of Mobile Edge Computing Network.}\label{fig:mec}
\end{figure*}

As depicted in Fig.~\ref{fig:mec}, MEC servers are implemented directly at the BSs using generic-computing platform, allowing the execution of applications in close proximity to end users. With this position, MEC can help fulfill the stringent low-latency requirement of 5G networks. Additionally, MEC offers various network improvements, including: (i) optimization of mobile resources by hosting compute-intensive applications at the network edge, (ii) pre-processing of large data before sending it (or some extracted features) to the cloud, and (iii) context-aware services with the help of RAN information such as cell load, user location, and allocated bandwidth. Although MEC principle also aligns with the concept of \emph{fog computing}~\cite{bonomi2012fog} and the two are often referred to interchangeably, they slightly differ from each other. While fog computing is a general term that opposes with cloud computing in bringing the processing and storage resources to the lower layers, MEC specifically aims at extending these capabilities to the edge of the RAN with a new function splitting and a new interface between the BSs and upper layer. Fog computing is most commonly seen in enterprise-owned gateway devices whereas MEC infrastructure is implemented and owned by the network operators.

Fueled with the potential capabilities of MEC, we propose a real-time context-aware collaboration framework that lies at the edge of the cellular network and works side-by-side with the underlying communication network. In particular, we aim at exploring the synergies among connected entities in the MEC network to form a heterogeneous computing and storage resource pool. To illustrate the benefits and applicability of MEC collaboration in 5G networks, we present three use- cases including mobile-edge orchestration, collaborative video caching and processing, and multi-layer interference cancellation. These initial target scenarios can be used as the basis for the formulation of a number of specific applications.

The remainder of this article is organized as follows: In Sect.~II, we present the state of the art on MEC; in Sect.~III, we provide a comparison between MEC and C-RAN in various features; in Sects.~IV, V, and VI, we describe the three case studies to illustrate the applicability and benefits of collaborative MEC paradigm; in Sect.~VII, we highlight some key challenges and open-research issues that need to be tackled; finally, we draw our conclusions in Sect.~VIII.

\section{State of the Art} \label{sec:state_art}
In 2013, Nokia Networks introduced a very first real-world MEC platform~\cite{nokiaRACS} in which the computing platform---Radio Applications Cloud Servers~(RACS)---is fully integrated with the Flexi Multiradio base station. 
Under a scenario of a ``smarter city''~\cite{IBM_smartercity}, IBM discusses how operators can leverage the capabilities of mobile edge network-virtualization to deploy disruptive services for consumers and enterprises. Saguna also introduces their fully virtualized MEC platform, so called Open-RAN~\cite{sagunaMEC}, that can provide an open environment for running third-party MEC applications. Recently, the European Telecommunications Standards Institute~(ETSI) formed a MEC Industry Specifications Group~(ISG) in order to standardize and moderate the adoption of MEC within the RAN~\cite{hu2015mobile}.

From the theoretical perspective, the authors in~\cite{sardellitti2015joint} consider the computation offloading problem in a multi-cell MEC network, where a dense deployment of radio access points facilitates proximity high-bandwidth access to computational resources but also increases inter-cell interference. The authors in~\cite{chiangFog} provide a collective overview of the opportunities and challenges of ``Fog computing'' in the networking context of the Internet of Things~(IoT). Several case studies are presented to highlight the potential and challenges of the Fog control plane such as interference, control, configuration, and management of networks, etc.\footnote{Refer to: http://Fogresearch.org}

In summary, prior works on MEC focused on feasibility of MEC-RAN integration, deployment scenarios, and potential services and applications. \emph{In contrast to existing works on MEC, which do not explore the synergies among the MEC entities, this article takes one step further by proposing a collaborative MEC paradigm and presents three strong use cases to efficiently leverage this collaboration space.}

\section{MEC versus C-RAN} \label{sec:mec_vs_cran}

\begin{table*}[t!]
\caption{Comparison of features: MEC versus C-RAN.}\centering
\label{table:MEC_CRAN}
\begin{center}
\begin{tabularx}{\linewidth}{|c|X|X|}
\hline  \\ [-1.5ex]
& \textbf{MEC} & \textbf{C-RAN} \\
\hline
\bottomrule \hline
\textbf{Location} &Co-located with base stations or aggregation points. & Centralized, remote data centers. \\[.5ex]

\hline
\textbf{Deployment Planning} &Minimal planning with possible ad-hoc deployments. &Sophisticated. \\[.5ex]

\hline
\textbf{Hardware} &Small, heterogeneous nodes with moderate computing resources. &Highly-capable computing servers. \\[.5ex]

\hline
\textbf{Front-haul Requirements} &Front-haul network bandwidth requirements grow with the total amount of data that need to be sent to the core network after being filter/processed by MEC servers. &Front-haul network bandwidth requirements grow with the total aggregated amount of data generated by all users. \\[.5ex]

\hline
\textbf{Scalability} &High &Average, mostly due to expensive front-haul deployment. \\[.5ex]

\hline
\textbf{Application Delay} &Support time-critical applications that require latencies less than tens of milliseconds. &Support applications that can tolerate round-trip delays in the order of a few seconds or longer.\\[.5ex]

\hline
\textbf{Location Awareness} &Yes &N/A \\[.5ex]

\hline
\textbf{Real-time Mobility} &Yes &N/A \\[.5ex]

\hline
\end{tabularx}
\end{center}
\end{table*}

A redesigned, centralization of RAN is proposed as C-RAN where the physical-layer communication functionalities are decoupled from the distributed BSs and are consolidated in a virtualized central processing center. With its centralized nature, can be leveraged to address the capacity fluctuation problem and to increase system energy efficiency in mobile networks~\cite{pompili2016elastic}. Beside an approach to 5G standardization, C-RAN can provide new opportunities for IoT, opening up a new horizon of ubiquitous sensing, interconnection of devices, service sharing, and provisioning to support better communication and collaboration among people and things in a more distributed and dynamic manner. The integration of cloud provider, edge gateways, and end-devices can support powerful processing and storage facilities to massive IoT data streams (big data) beyond the capability of individual ``things” as well as provide automated decision making in real time. Thus, the C-RAN and IoT convergence can enable the development of new innovative applications in various emerging areas such as smart cities, smart grids, smart healthcare, and others aimed at improving all aspects of human life.

The full centralization principle of C-RAN, however, entails the exchange of radio signals between the radio heads and cloud processing unit, which imposes stringent requirement to the fronthaul connections in terms of throughput and latency. On the other hand, MEC paradigm is useful in reducing latency and improving localized user experience, but the amount of processing power and storage is orders of magnitude below that of the centralized cloud in C-RAN. In Table~\ref{table:MEC_CRAN}, we summarize the comparison between MEC and C-RAN in various aspects. One important note is that MEC does not contradict with C-RANs but rather complement them. For example, an application that needs to support very low end-to-end delay can have one component running in the MEC cloud and other components running in the distant cloud.

In the following sections, we present our case studies where we propose novel scenarios and techniques to take advantage of the collaborative MEC systems.

\section{Case Study I: Mobile Edge Orchestration} \label{sec:computing}

In spite of the limited resources (e.g., battery, CPU, memory) on mobile devices, many computation-intensive applications from various domains such as computer vision, machine learning, and artificial intelligence are expected to work seamlessly with {\it real-time} responses. However, the traditional way of offloading computation to the remote cloud often leads to unacceptable delay (e.g., hundreds of $\rm{ms}$~\cite{Cuervo2010}) and heavy backhaul usage. Owing to its distributed computing environment, MEC can be leveraged to deploy applications and services as well as to store and process content in close proximity to mobile users. This would enable applications to be split into small tasks with some of the tasks performed at the local or regional clouds as long as the latency and accuracy are preserved.

\begin{figure*}[t]
 \centering
 \begin{tabular}{ccc}
\hspace*{-.5cm}\includegraphics[scale = .4]{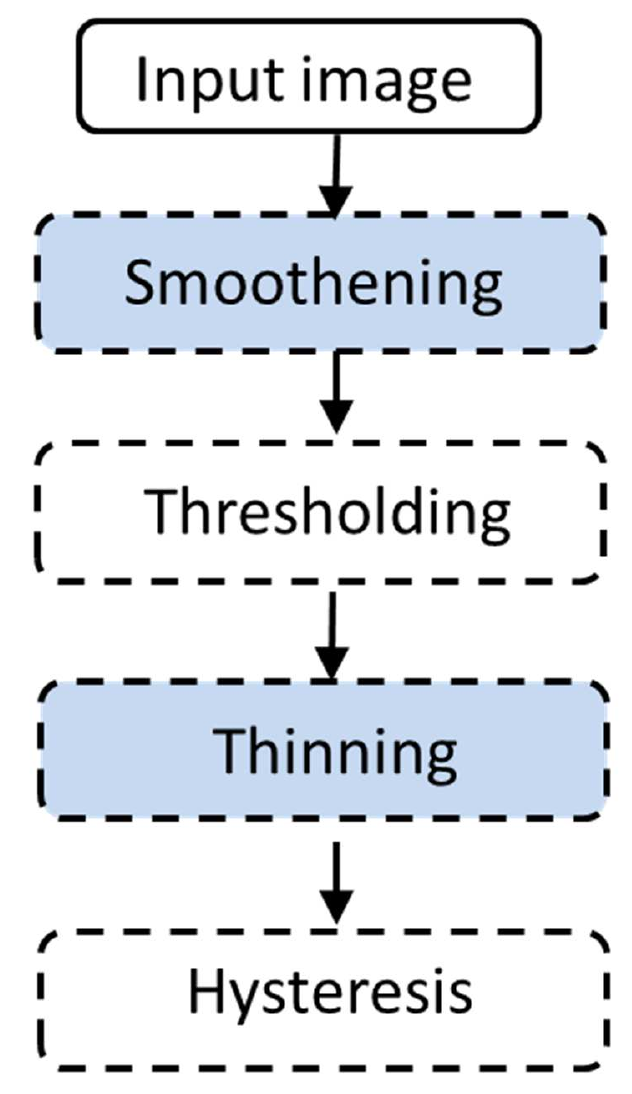} &
\hspace*{-.0cm}\includegraphics[scale = .5]{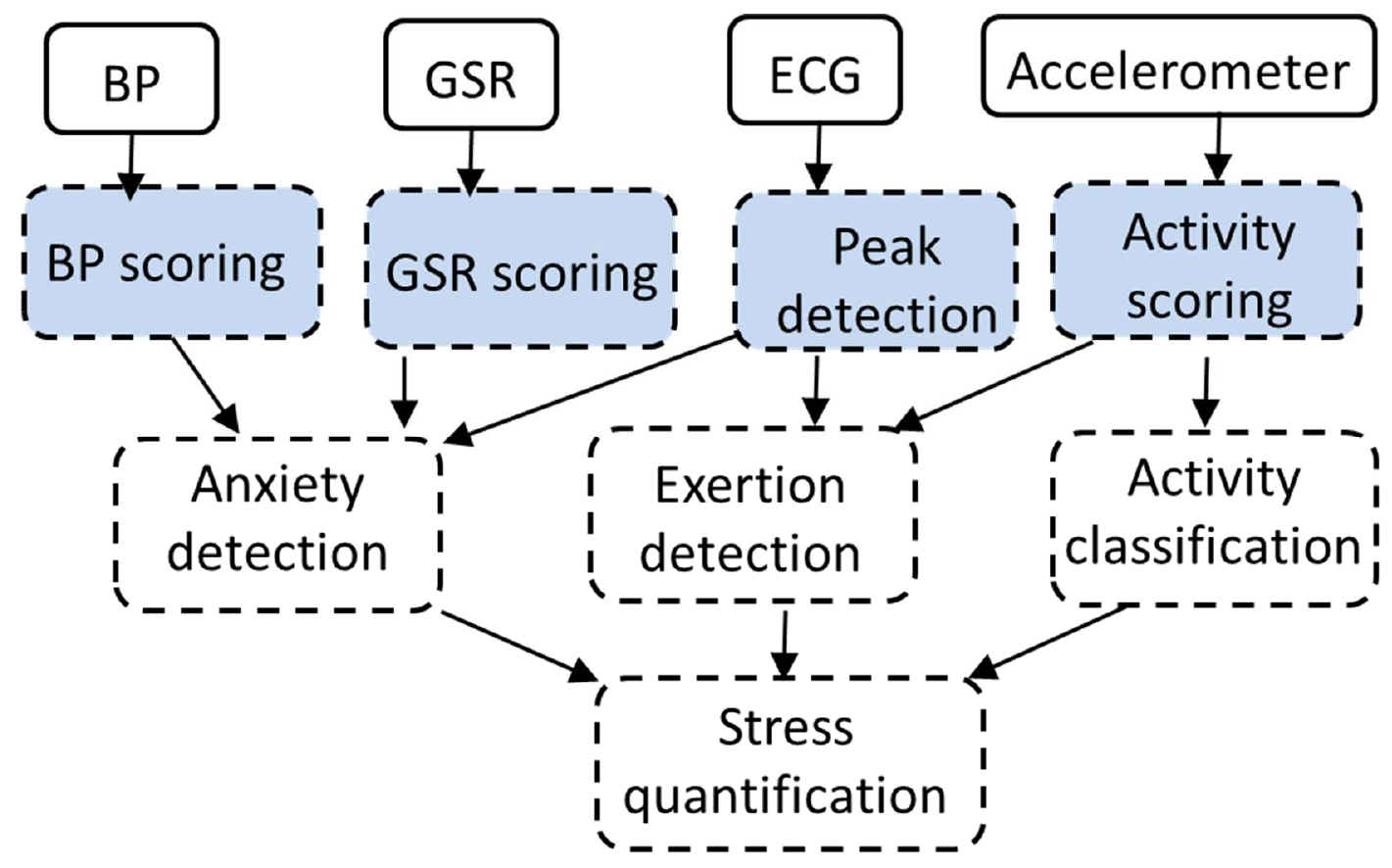} &
\hspace*{-.0cm}\includegraphics[scale = .5]{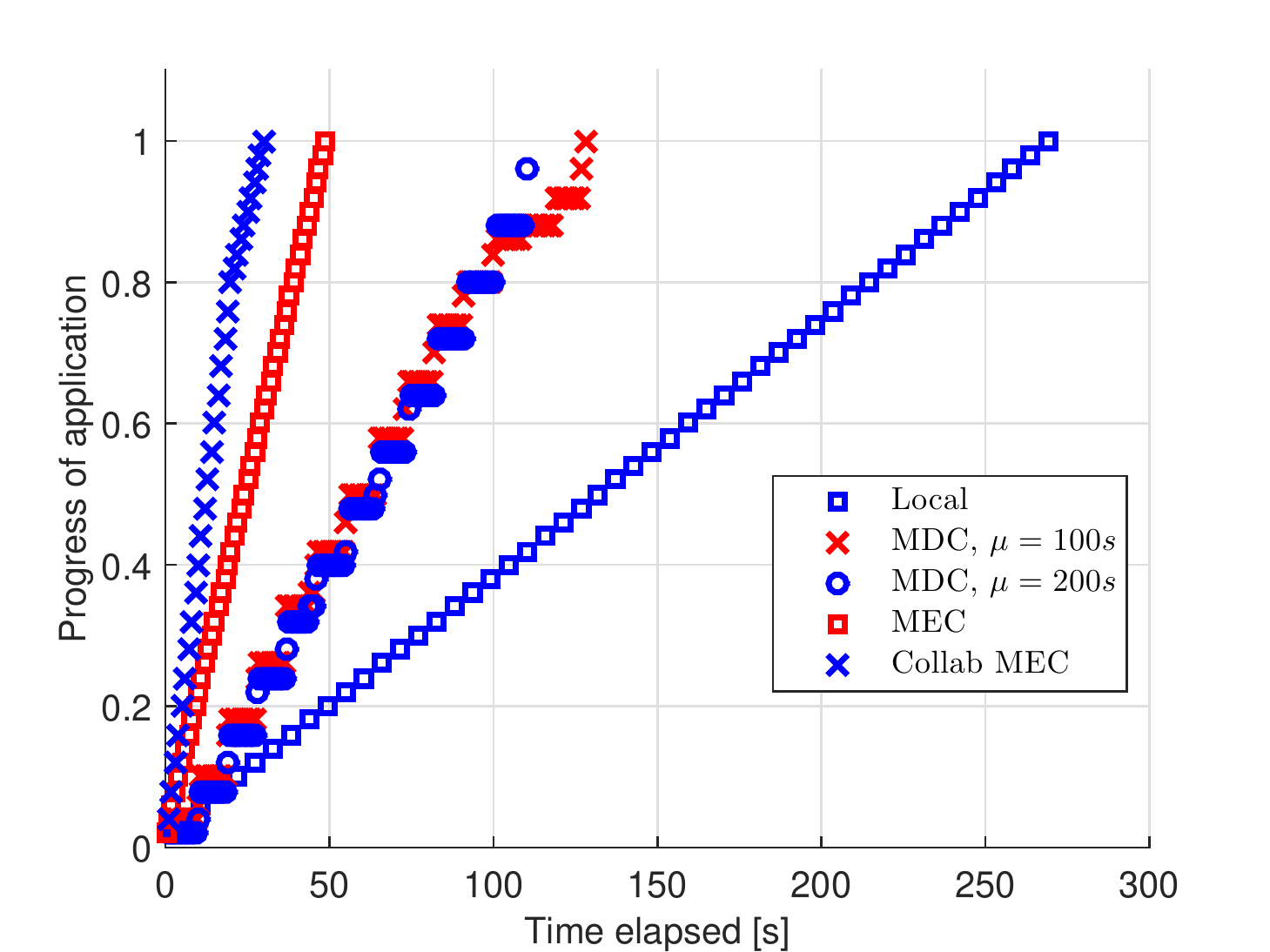} \\
 \small(a) & \small(b) & \small(c)
\end{tabular}
\caption{Block diagram showing tasks of different mobile applications in (a) image processing domain (Canny-edge detection); (b) ubiquitous health-care domain (Stress quantification). The blue blocks in these applications represent the computationally-intensive tasks of the applications that can be offloaded to the remote resources (edge and cloud); (c) Comparison of different startegies to execute computationally-intensive mobile applications.}\label{fig:MECrowdComp}
\end{figure*}

\emph{In this case study, we envision a collaborative distributed-computing framework where resource-constrained end-user devices outsource their computation to the upper-layer computing resources at the edge and cloud layers.} Our framework extends the standard MEC originally formulated by ETSI, which only focuses on individual MEC entities and on the vertical interaction between end-users and a single MEC node. conversely, our proposed collaborative framework will bring many individual entities and infrastructures to collaborate with each other in a distributed system. In particular, our framework oversees a hierarchical architecture consisting of: i) \textit{end-user}, which implies both mobile---and static---end-user devices such as smart phones, sensors, actuators, ii) \textit{edge nodes}, which are the MEC servers co-located with the BSs, and iii) \emph{cloud node}, which is the traditional cloud-computing server in a remote datacenter. Our novel resource-management framework lies at the intermediate edge layer and orchestrates both the \emph{horizontal} collaboration at the end-user layer and the MEC layer as well as the \emph{vertical} collaboration between end-users, edge nodes, and cloud nodes. The framework will make dynamic decisions on ``what'' and ``where'' the tasks in an application should be executed based on the execution deadline, network conditions, and device battery capacity.

There have been a number of works in mobile computing domain where data from the local device is uploaded to the cloud for further processing~\cite{Gordon2012} or executed locally via approximate computing~\cite{pandey2017exploiting} to combat the problem of limited resources. In~\cite{CumulusDAS16} we focused on the ``extreme'' scenario in which the resource pool was composed purely of proximal mobile devices. In contrast, MEC introduces a new stage of processing such that the edge nodes can analyze the data from nearby end-users and notify cloud node for further processing only when there is a significant change in data or accuracy of results. In addition, sending raw-sensor values from end-users to the edge layer can overwhelm the fronthaul links, hence, depending on the storage and compute capabilities of user devices and the network conditions, the MEC orchestrator can direct the end-users to extract features from the raw-data before sending to the edge nodes. 

In Figs.~\ref{fig:MECrowdComp}(a, b), we illustrate two mobile applications from different domains that are good candidates of being executed at the edge. The blue blocks in these applications represent the computation-intensive tasks of the applications that can be offloaded to the upper-level resources (edge and cloud). In Fig.~\ref{fig:MECrowdComp}(c) we compare the time taken for execution of the mobile application represented in Fig.~\ref{fig:MECrowdComp}(a) (Canny-edge detection) by using different strategies: (i) executing the application locally on the mobile device (\emph{Local}), (ii) distributing tasks to proximal mobile devices forming a Mobile Device Cloud~(\emph{MDC})~\cite{CumulusDAS16}, (iii)-(iv) offloading the tasks to a single MEC server (\emph{MEC}), and to two collaborating MEC servers (\emph{Collab MEC}), respectively. For execution in an MDC we model the mobility patterns of devices in the proximity as a normal distribution with mean availability duration of devices varying with $\mu=\{\mathrm{100}, \mathrm{200}\}~\mathrm{s}$ and $\sigma$=$\mathrm{5}~\mathrm{s}$. We assume the local mobile devices connect with the MEC server on a $\mathrm{1}~\mathrm{Mbps}$ link. The mobile devices involved in the experiment include 2 Samsung Galaxy Tabs and 4 smartphones (2 ZTE Avid N9120's and 2 Huawei M931's). For MEC servers we used two desktops with Intel Core i7 CPU at $\mathrm{3.40}$~GHz  and $\mathrm{16}$~GB RAM. We execute the application in Fig.~\ref{fig:MECrowdComp}(a) by using input data from the Berkeley image segmentation and benchmark dataset. Resolution of each image is $\mathrm{481} \times \mathrm{321}$ pixels. A task consists of finding edges of 20 images from the dataset. For the current simulation, we use a round-robin technique for the MDC where all the devices are given equal tasks. Sophisticated task-allocation algorithms can be run at the arbitrator to decide how many tasks to run at each service provider based on the computational capabilities of different service providers. After execution of the tasks, the service provider returns the task to the service requester.
In Fig.~\ref{fig:MECrowdComp}(c) we see that the performance of execution on a single MEC server is significantly better than the execution on a local device and MDC. The gain in terms of execution time on using collaborative MEC over execution of the application on a single MEC server is around 40\%. 

The example above illustrates the benefit of collaborative MEC framework in reducing execution time of the two image processing tasks. The extension of such strategy will greatly benefit the service requesters, which are the health analytics providers in this case, as they see lower latency in the execution of the application as the MEC servers are at the BS rather than at the Cloud. These service requesters require processing of large data and the MEC servers expedite the processing time by dividing the processing between MEC servers (extracting features from the raw data) and cloud resources (running computation-intensive applications using extracted features as input data). This leads to faster availability results for the data analytics expert and also gives faster result to patients requesting results.

Currently, to present preliminary results we use a simple image processing application. However, we believe that a compute-intensive application (such as real-time activity detection with significant variations in execution time of tasks) or a data-intensive application (such as real-time face-detection in a video with large volume of input data) will require a powerful computing environment as ours  to make dynamic decisions of what and where the tasks to be executed based on real-time conditions, which will make application execution via collaborative MEC even more challenging.

\section{Case Study II: Collaborative Video Caching and Processing} \label{sec:caching}

\begin{figure}
 \centering
 \includegraphics[width=0.45\textwidth]{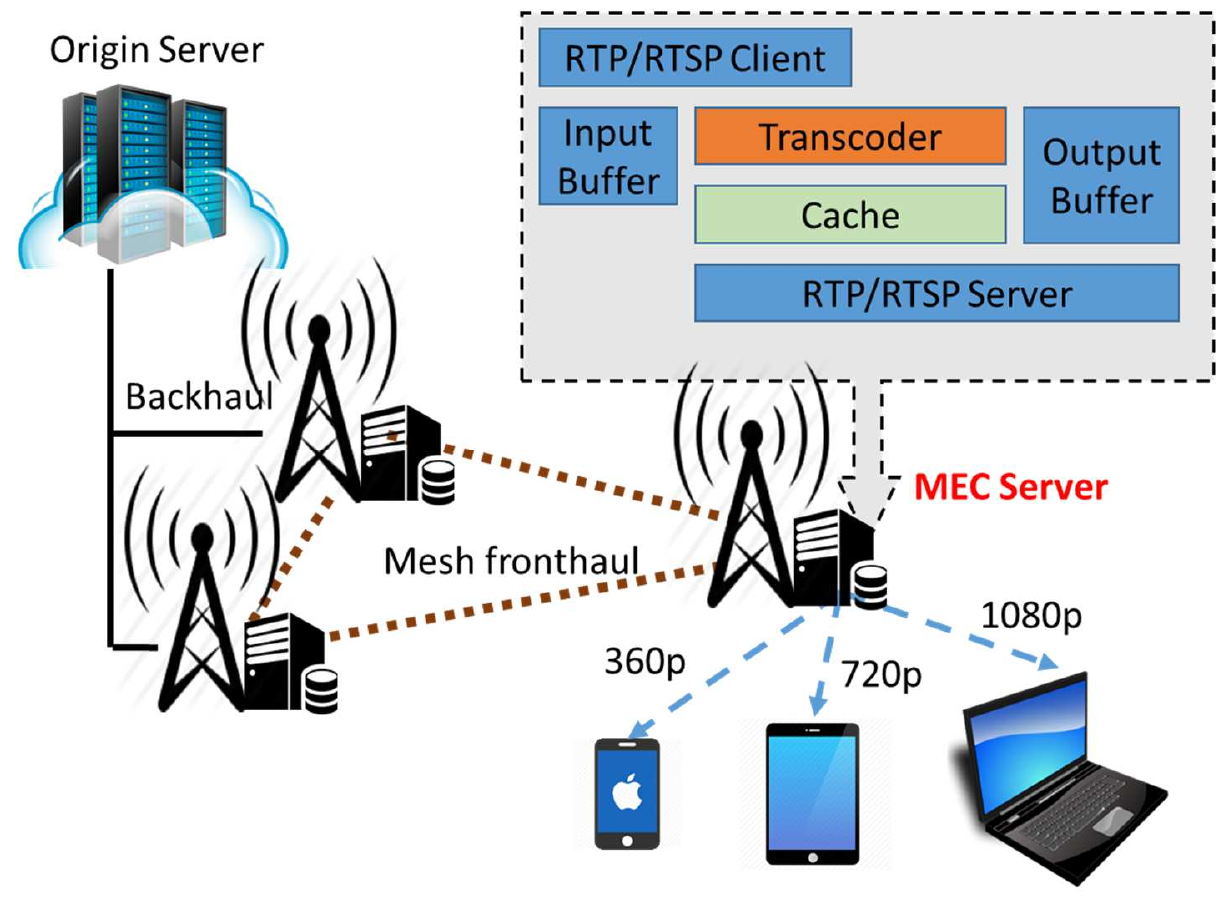}
\caption{Illustration of collaborative video caching and processing framework deployed on MEC network.}\label{fig:mec_cache}
\end{figure}

Mobile video streaming traffic is predicted to account for 72$\%$ of the overall mobile data traffic by 2019\footnote{Refer to ``Global Mobile Data Traffic Forecast Update 2014--2019. White~Paper~c11-520862" by Cisco Visual Networking Index.}, posing immense pressure on network operators. To overcome this challenge, edge caching has been recognized as a promising solution, by which popular videos are cached in the BSs or access points so that demands from users to the same content can be accommodated easily without duplicate transmission from remote servers. This approach helps substantially reduce backhaul usage and content access delay. While content caching and delivery techniques in wireless networks have been studied extensively (see, e.g.,~\cite{bastug2014living} and references therein), existing approaches rarely exploit the synergy of caching and computing at the cache nodes. Due to the limited cache storage at individual BSs, the cache hit rate is still moderate. Several solutions have considered collaborative caching, in which a video request can be served using not only the local BS`s cache, but also the cached copy at neighboring BSs via the backhaul links~\cite{tran2016octopus}.

\begin{figure*}[t]
 \centering
 \begin{tabular}{ccc}
\hspace*{-.3cm}\includegraphics[scale = .6]{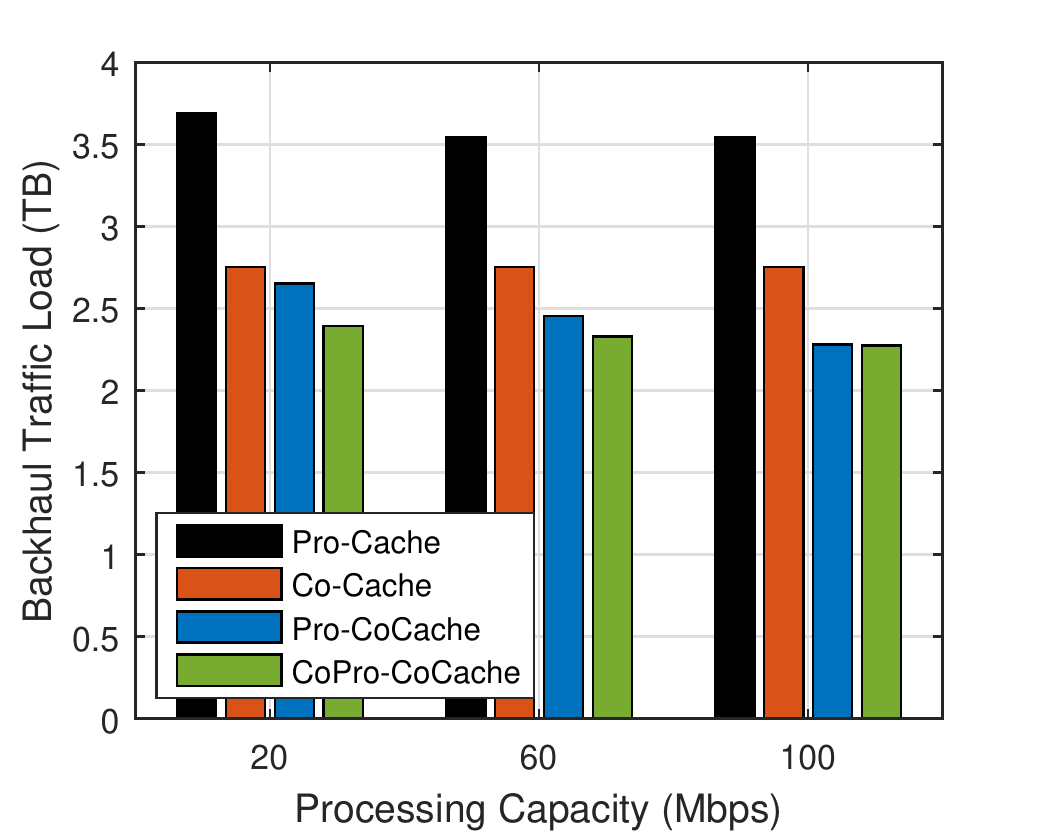} &
\hspace*{-.6cm}\includegraphics[scale = .6]{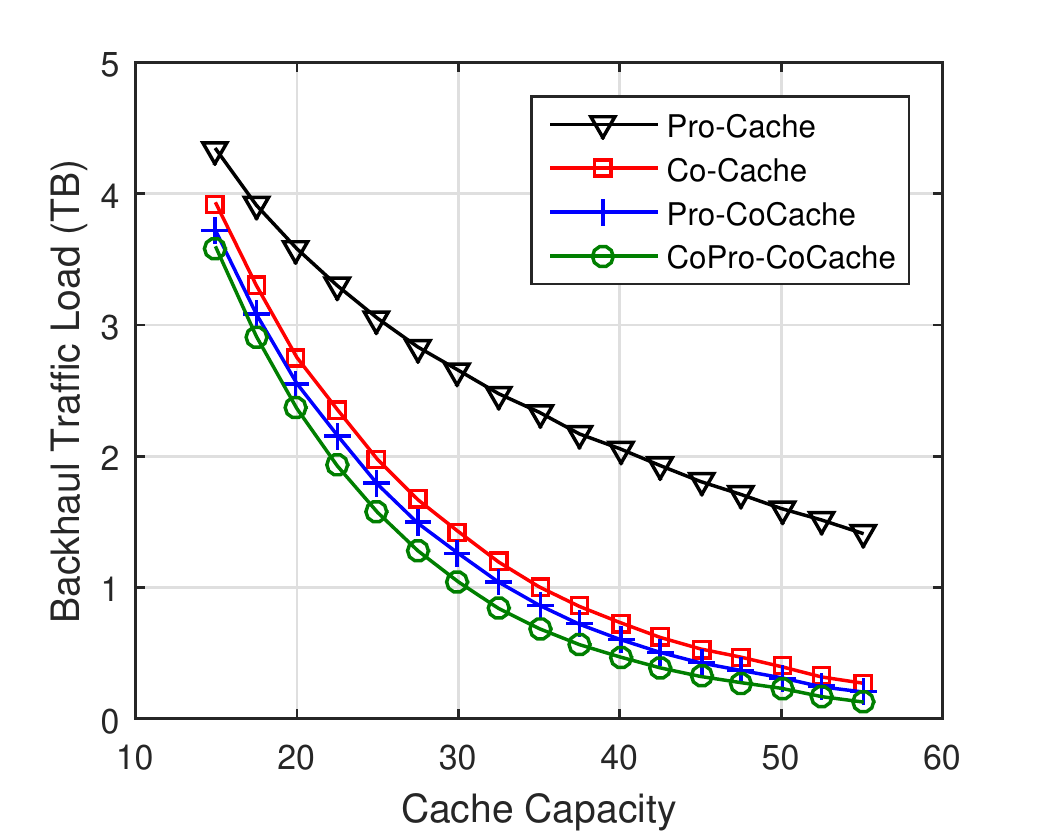} &
\hspace*{-.6cm}\includegraphics[scale = .6]{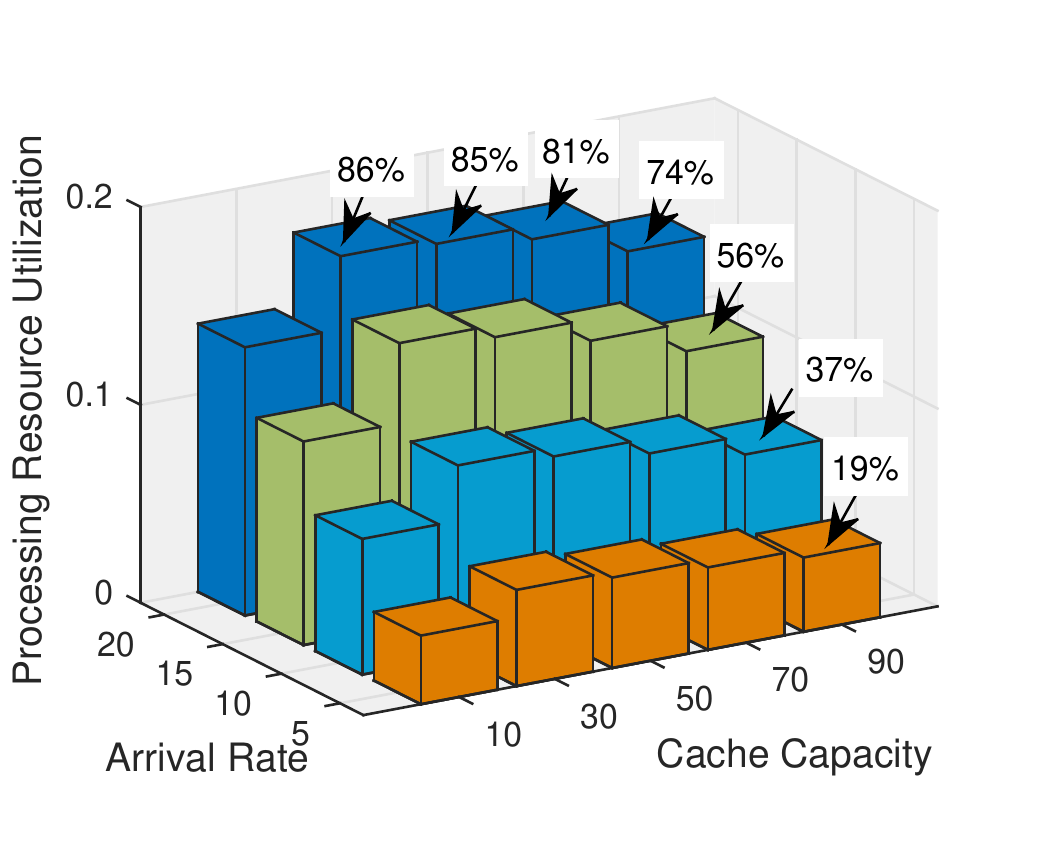} \\
 \small(a) & \small(b) & \small(c)
\end{tabular}
\caption{ Considered caching strategies: (i) \emph{Pro-Cache}---non-collaborative caching with processing, (ii) \emph{Co-Cache}---collaborative caching without processing, (iii) \emph{Pro-CoCache}---collaborative caching with processing and (iv) \emph{CoPro-CoCache}---collaborative caching with collaborative processing (\emph{proposed}); Video duration is set to 10 \rm{min}, and each video has four variants with relative bit rates of 0.82, 0.67, 0.55 and 0.45 of the original video bit rate ($2~\mathrm{Mbps}$); (a) Backhaul traffic load versus Processing Capacity ($\mathrm{Mbps}$) with Cache Capacity = $30\%$ library size, (b) Backhaul traffic load versus Cache Capacity, (c) Processing resource utilization versus Arrival Rate (request/BS/min) and Cache Capacity, in (b, c) we set Processing Capacity = $40~\mathrm{Mbps}$.
}\label{fig:cache_replacement}
\end{figure*}

\emph{With the emergence of MEC, it is possible to not only perform edge caching but also edge processing. Our approach will leverage edge processing capability to improve caching performance/efficiency. Such joint caching and processing solution will trade off storage and computing resources with backhaul bandwidth consumption, which directly translates into sizable network cost saving}. Due to the heterogeneity of users' processing capabilities and the varying of network connections, user preference and demand towards a specific video might be different. For example, users with highly capable device and fast network connection usually prefer high resolution videos whereas users with low processing capability or low bandwidth connection may not enjoy high quality videos because the delay is large and the video may not fit within the device's display. Leveraging such behavior, Adaptive Bit Rate~(ABR)\footnote{Refer to: https://en.wikipedia.org/wiki/\text{Adaptive\_bitrate\_streaming}} streaming techniques have been developed to improve the quality of delivered video on the Internet as well as wireless networks. Examples of such techniques include Apple HTTP Live Streaming~(HLS), Microsoft Smooth Streaming and Adobe Systems HTTP Dynamic Streaming. In ABR streaming, the quality of the streaming video is adjusted according to the user device's capabilities, network connection and specific request. Existing video caching systems often treat each request for a video version equally and independently, without considering their transcoding relationship, resulting in moderate benefits.

In this case study, we exploit both ABR streaming and collaborative caching to improve the caching benefits beyond what can be achieved by traditional approaches. The proposed collaborative video caching and processing framework deployed on MEC network~\cite{tran2017wons} is illustrated in Fig.~\ref{fig:mec_cache}. Given the storage and computing capabilities, each MEC server acts as a cache server and also a transcoding server. These servers collaborate with each other to not only provide the requested video but also transcode it to an appropriate variant. Each variant is a bit-rate version of the video and a higher bit-rate version can be transcoded into a lower bit-rate ones. The potential benefits of this strategy is three-fold: (i) the content origin servers need not generate all variants of the same video, (ii) users with various capabilities and network conditions will receive videos that are suited for their capabilities, as content adaptation is more appropriately done at the network edge, and (iii) collaboration among the MEC servers enhances cache hit ratio and balance processing load in the network.

In our proposed joint collaborative caching and processing strategy,  referred to as \emph{CoPro-CoCache}, we distribute the most popular videos in the serving cell of each BS to the corresponding cache server of that BS, until the cache storage is full. When a user requests for a video that requires transcoding from a different version in the cache, the transcoding task is assigned to the MEC server having lower load, which could be the MEC server storing the original video version (data provider node) or the serving MEC server (delivery node). This helps balance the processing load in the network.

To illustrate the potential benefits of the proposed approach, we perform numerical simulation on a representative RAN with 5 BSs, each equipped with a MEC server that performs caching and transcoding. We assume a library of $1000$ videos is available for download. The video popularity requested at each BS follows a Zipf distribution with parameter $0.8$, i.e., the probability that an incoming request is for the $i$-th most popular video is proportional to $1/i^{0.8}$. In order to obtain a scenario where the same video can have different popularities at different locations, we randomly shuffle the distributions at different BSs. Video request arrival follows a Poisson distribution with same rate at each BS. In Fig.~\ref{fig:cache_replacement}(a, b) we compare the performance of four caching strategies in terms of backhaul traffic reduction. It can be seen that utilizing processing capabilities significantly helps reducing the backhaul traffic load. In addition, our proposed \emph{CoPro-CoCache} strategy explores the synergies of processing capabilities among the MEC servers, rendering additional performance gain. Fig.~\ref{fig:cache_replacement}(c) illustrates the processing resource utilization of \emph{CoPro-CoCache} scheme versus different video request arrival rates and cache capacity. We observe that the processing utilization increases with arrival rate and moderate cache capacity, however it decreases at high cache capacity. This is because with high cache capacity, we can store almost all the popular videos and their variants and thus there are fewer requests requiring transcoding.

While choosing the optimal bitrate for video streaming can enhance instant download throughput, existing client-based bitrate selection may not be able to adapt fast enough to the rapidly varying conditions, leading to under-utilization of radio resources and suboptimal user experience. A promising solution is to use a RAN analytic agent at the MEC server to inform the video server of the optimal bitrate to use given the radio conditions for a particular video request from end-user. Designing an efficient solution to address bitrate adaption with respect to channel conditions is still an open problem.

\section{Case Study III: Two-Layer Interference Cancellation} \label{sec:interference}
Deploying more small-cell BSs can improve spectral efficiency in cellular network, however making inter-cell interference become more prominent. To mitigate such interference, a promising approach is to employ Coordinated MultiPoint~(CoMP) transmission and reception techniques. In CoMP, a set of neighboring cells are divided into clusters; within each cluster, the BSs are connected with each other via a fixed Backhaul Processing Unit~(BPU) and exchange Channel State Information~(CSI) as well as Mobile Station~(MS) signals to cancel the intra-cluster interference. However, CoMP does not take into account the inter-cluster interference, resulting in moderate improvement in system capacity. Furthermore, the additional processing required for multi-site reception/transmission, CSI acquisition, and signaling exchanges among different BSs could add considerable delay and thus limit the cluster size in order to comply with the stringent delay requirement in 5G networks. In addition, applying CoMP for all users might be unnecessary as certain users, especially those at the cell centers, often have high level of Signal-to-Interference-plus-Noise-Ratio~(SINR) and do not cause intense interference to the neighboring BSs. 

\emph{To overcome the existing challenges of CoMP and reduce the latency and bandwidth between the BSs and the BPU, we advocate a two-layer interference cancellation strategy for an uplink MEC-assisted RAN}. In particular, based on the Channel Quality Indicator~(CQI) of each user, our solution identifies ``where'' to process its uplink signal so as to reduce complexity, delay and bandwidth usage. In a MEC-assisted RAN, we have access to the computational processing at the BSs, the signal demodulation of the cell-center MSs can be done in local BSs (layer~1). This means that the system performance for cell-center MSs relies on simple single transmitter and receiver. On the other hand, since the SINR of cell-edge MSs are often low, their signals should be transmitted to the BPU (layer~2) for further processing. In this case, the BPU has access to all the cell-edge MSs from different cells and is able to improve their SINRs via coordinated processing. 

\begin{figure}
 \centering
 \includegraphics[width=0.35\textwidth]{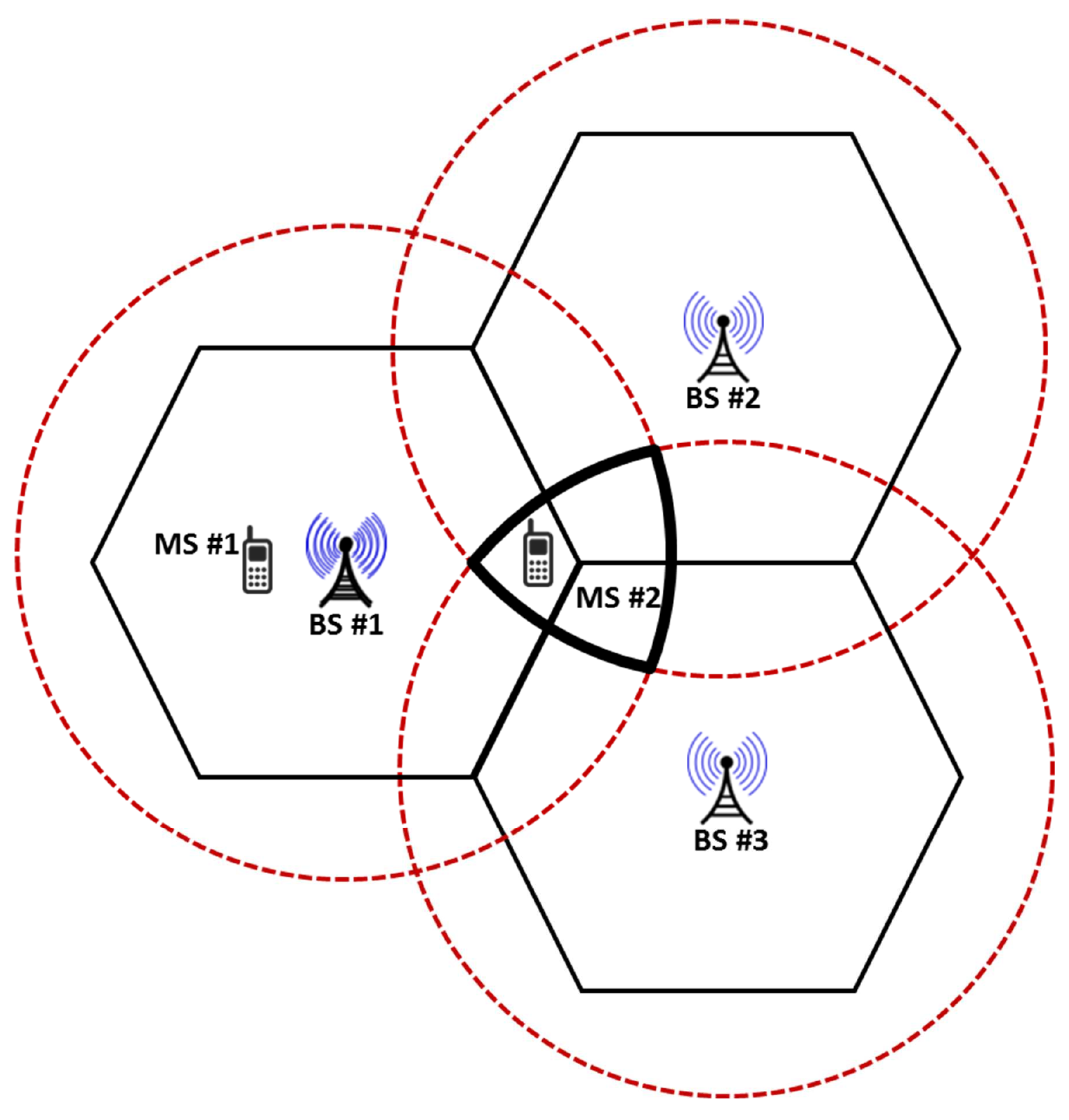}
\caption{Mobile Station (MS) \#1 and \#2 are located at cell-center and cell-edge regions, respectively. Since the MS \#1 is far from the neighbouring BSs, the signal demodulation can be performed at the edge (layer 1). However, MS \#2 is located at the cell-edge region and its interference to the neighbouring BSs should be canceled at upper layer (layer 2).}\label{fig:2IC}
\end{figure}

As illustrated in Fig.~\ref{fig:2IC}, each red dotted circle indicates the interference region of the corresponding cell which is defined as a region within which if MSs from other cells moved in, they could render an ``intense'' interference at the BS serving the cell. Since MS \#1 is a cell-center MS and is outside the interference region of BSs \#2 and \#3, its interference at those BSs is low due to the high path-loss; hence, there is no need to employ coordinated interference cancellation for MS \#1 and thus its signal demodulation can be performed at the edge layer. Conversely, since MS \#2 is a cell-edge MS and is located in the interference region of BSs \#2 and \#3, there may be an intense interference from MS \#2 to BSs \#2 and \#3; thus, coordinated interference cancellation at the upper layer is needed to cancel this interference and the BS should transmit the raw data to the upper layer for further processing.

\section{Challenges and Open-Research Issues} \label{sec:challenges}
The decentralization of cloud computing infrastructure to the edge brings various benefits that contribute to the 5G evolution, while at the same time introduces new challenges and open-research issues as highlighted in the following.

{\bf{Resource management.}} The computing and storage resources in individual MEC platform are expected to be limited and may be able to support a constrained number of applications with moderate needs of such resources. Currently, network providers often race for extensively stand-alone infrastructures to keep up with the demand while struggling with lower return on investment. An alternative approaches such as MEC as a Service may need to be considered, whereby operators' resources can be opened up for interested service providers to request or relinquish based on service demand.

{\bf{Interoperability.}} MEC infrastructures owned by different network providers should be able to collaborate with each other as well. This necessitates the specification of common collaboration protocols, also allowing for service providers to access network and context information regardless of their deployment place.

{\bf{Service discovery.}} Exploiting the synergies of distributed resources and various entities, as envisioned in our mobile edge orchestration framework, requires discovery mechanisms to find appropriate nodes that can be leveraged in a decentralized set up. Automatic monitoring of the heterogeneous resources and accurate synchronization across multiple devices are also of great importance.

 {\bf{Mobility support.}} In a small cell network the range of each individual cell is limited. Mobility support becomes more important and solution for a fast process migration may become necessary.

 {\bf{Fairness.}} Ensuring fair resource sharing and load balancing is also an essential problem. There is potential that a small number of nodes could carry the burden of processing, while a large number of nodes would contribute little to the efficiency of the distributed network.
 
{\bf{Security.}} Security issues might hinder the success of MEC paradigm if not carefully considered. Existing centralized-authentication protocols might not be applicable for some parts of the infrastructure that have limited connectivity to the central authentication server. It is also important to implement trust management systems that are able to exchange compatible trust information with each other, even if they belong to different trust domains. Furthermore, as service providers want to acquire user information to tailor their services (e.g., content providers want to know users’ preference and mobility patterns to cache proactively their contents, as discussed in case study II), there is a great challenge to the development of privacy-protection mechanisms that can efficiently protect users’ location and service usage.  
\section{Conclusions} \label{sec:conclusion}
Mobile-Edge Computing (MEC) enables a capillary distribution of cloud computing capabilities to the edge of the radio access network. This emerging paradigm allows for execution of delay-sensitive and context-aware applications in close proximity to the end-users while alleviating backhaul utilization and computation at the core network. This article proposes to explore the synergies among connected entities in the MEC network to form a heterogeneous resource pool. We present three representative use-cases to illustrate the benefits of MEC collaboration in 5G networks. Technical challenges and open-research issues are highlighted to give a glimpse idea on the development and standardization roadmap of mobile-edge ecosystem.

\balance
\small



\vspace{0.4cm}
\section*{Biography}\label{sec:bio}

\vspace{0.2cm}
\textbf{Tuyen X. Tran} is working towards his PhD degree in Electrical and Computer Engineering~(ECE) at Rutgers U. under the guidance of Dr.~Pompili. He received the MSc degree in ECE from the U. of Akron, USA, in 2013, and the BEng degree (Honors Program) in Electronics and Telecommunications from Hanoi U. of Technology, Vietnam, in 2011. His research interests are on the application of optimization, statistics, and game theory to wireless communications and cloud computing.

\vspace{0.2cm}
\textbf{Abolfazl Hajisami} started his PhD program in ECE at Rutgers U. in 2012. Currently, he is pursuing research in the fields of C-RAN, cellular networking, and mobility management under the guidance of Dr.~Pompili. Previously, he had received his MS and BS from Sharif University of Technology and Shahid Beheshti University (Tehran, Iran), in 2010 and 2008, respectively. His research interests are wireless communications, cloud radio access network, statistical signal processing, and image processing.

\vspace{0.2cm}
\textbf{Parul Pandey} is a PhD candidate in the Dept. of ECE at Rutgers U. She is currently working on mobile and approximate computing, cloud-assisted robotics, and underwater acoustic communications under the guidance of Dr.~Pompili as a member of the CPS-Lab. Previously, she had received her BS degree in electronics and communication engineering from Indira Gandhi Institute of Technology,
Delhi, India and her MS degree in ECE from the U. of Utah, USA, in 2008 and 2011, respectively.

\vspace{0.2cm}
\textbf{Dario Pompili} is an Assoc. Prof. with the Dept. of ECE at Rutgers U., where he directs the Cyber-Physical Systems Laboratory~(CPS-Lab). He received his PhD in ECE from the Georgia Institute of Technology in 2007. He had previously received his `Laurea' (combined BS and MS) and Doctorate degrees in Telecommunications and Systems Engineering from the U. of Rome ``La Sapienza,'' Italy, in 2001 and 2004, respectively. He is a recipient of the NSF CAREER'11, ONR Young Investigator Program'12, and DARPA Young Faculty'12 awards. He is a Senior Member of both the IEEE Communications Society and the ACM.

\end{document}